# An Integrated Human-Computer System for Controlling Interstate Disputes

Tshilidzi Marwala[1], Monica Lagazio[2] and Thando Tettey[1]

[1]School of Electrical and Information Engineering

University of the Witwatersrand

Private Bag x3

Wits, 2050, Republic of South Africa

E-mail: t.marwala@ee.wits.ac.za

[2]Department of Politics and International Relations

University of Kent

Rutherford College

Canterbury, Kent

CT2 7NX

United Kingdom

E-mail: m.lagazio@kent.ac.uk

**Abstract**

In this paper we develop a scientific approach to control inter-country conflict. This system makes use of a neural network and a feedback control approach. It was found that by controlling the four controllable inputs: Democracy, Dependency, Allies and Capability simultaneously, all the predicted dispute outcomes could be avoided. Furthermore, it was observed that controlling a single input Dependency or Capability also avoids all the predicted conflicts. When the influence of each input variable on conflict is assessed, Dependency, Capability, and Democracy emerge as key variables that influence conflict.

1. Introduction

With the frequency at which wars are occurring, it has become imperative that more research effort be directed towards conflict management. The main aim of conducting this research is to better understand the occurrence and management of International conflict. A significant amount of the research effort has been channeled towards conducting empirical studies of International conflict. These empirical studies have been advanced on two fronts. Firstly, there has been significant effort dedicated to improving the explanatory variable of interstate interactions. Secondly there has been an effort in finding a suitable model which will allow for the accurate forecasting of international conflict. A successful outcome in the quantitative analysis has therefore been defined as the ability to accurately forecast international conflict and at the same time give a causal explanation of dispute outcomes[1-7]. This result can then be specified as a tool which will contribute to decision making and policy formulation.

Recent developments on the data collection have allowed a significant improvement in the prediction of Militarized Interstate Disputes (MIDs). In this paper, MIDs are defined as disputes between sovereign states below the threshold of war and include: explicit threats to use force, display of force, mobilization of force, or the use of force short of war. This allows studies to include a broad scope of conflicts that pose a grave threat to international peace and security and have political effects and complications comparable to that of full warfare. On the forecasting side, there is a need to find more accurate ways of predicting international conflict. This has seen a shift from statistical techniques to neural networks in an attempt to avoid the problems experienced with statistical models. Neural networks have the disadvantage of not being transparent and therefore are not able to readily offer a causal explanation for results obtained. In the field of interstate conflict, no studies up to now have dealt with the issue of non-linear control for interstate disputes through the utilization of non-linear prediction models such as neural networks to build effective decision support systems for conflict prevention. Although some attempts have been made to use control theory in political science[9], these effort rely on traditional control mechanisms, which are still based on a linear plant model. However, we know that this is not an accurate method for conflict data since the relationships among key variables have been deemed to be highly interdependent and non-linear[2,3].

This paper implements the Bayesian framework to produce a neural network model which predicts MIDs. As a means of promoting transparency in the neural network model, automatic relevance determination (ARD)[10] is introduced, to understand the effects of the dyadic variables on MIDs sensitivity. Thirdly, this paper introduces neural network control approach as an infrastructure that can be used to control interstate disputes as well as further clarify the effects of dyadic variables. To achieve the initial goal, Bayesian neural networks are trained using the hybrid Monte Carlo (HMC) and Gaussian Approximation (GA) methods[10-12]. The neural networks developed are used to model the relationships between the dyadic attributes and the MIDs, and the HMC and GA results are then compared. The results and conclusions obtained from these investigations are then reported.

## 2. Background

As mentioned previously, Militarized interstate disputes (MIDs) are defined as a set of interactions between or among states that can result in the actual use, display or threat of using military force in an explicit way[1]. These interactions can result in either peace or conflict. The question of what explanatory variables determine MID is a current topic of research. Projects such as the Correlates of War (COW) facilitate the collection and dissemination and use of accurate and reliable quantitative data[8]. To further improve the quantitative study of international conflict, a generic term of "conflict" rather than "war" or "dispute" has been adopted. The use of MID data allows us not only to concentrate on intense state interactions but also on sub-war interactions, were militarized behavior occurs without escalation to war, as these may be very important in exploring mediation issues. On the forecasting side, statistical methods have been used for a long time to predict conflict and it has been found that no statistical model can predict international conflict with probability of more than 0.5 [3]. The use of statistical model has led to fragmentary conclusions and results that are not in unison. An example of this can be found in the investigation of the relationship between democracy and peace. Thomspon and Tucker[13], in their work, conclude that if the explanatory variables indicate that countries are democratic, the chances of war are then reduced. On the other hand Mansfield and Snyder[14] oppose this notion and suggest that democratization increases the likelihood of war. Lagazio and Russet[4] point out that the reason for the failure of statistical methods might be attributed to the fact that the interstate variables related to MID are non-linear, highly interdependent and context dependent. This therefore calls for the use of more suitable techniques. Neural networks, namely multi-layer perceptrons (MLPs), have been applied to the modeling of interstate conflict[3]. The advantage of using the MLP is that it models complex input-output relationships without the need for a priori knowledge or assumptions about the problem domain. The only problem with neural networks in this context is that the conflict data is skewed which has the effect of biasing the result towards more common events, giving a poor classification of the rare events. In this paper this is dealt with by training the neural network on a balanced dataset and then testing on an unbalanced test set.

## 3. Modeling of Conflict

*Modeling Data*

This section describes the variables and MID data that are used to construct the Bayesian neural network models. Our output is the MID, which represents the threat to use military force or a display of military force, which is conducted in an explicit and overtly non-accidental way between states[7]. This output is coded 1 if a militarized interstate dispute was begun and 0 otherwise. Only the initial year of the militarized conflict is included, since our concern is to predict the onset of a conflict rather than its continuation. For the inputs, we utilize the theoretical prospective and the five variables as described extensively and used by Russett and Oneal[1]. The first variable is *Allies*, a binary measure coded 1 if the members of a dyad are linked by any form of military alliance or 0 in the absence of military alliance. *Contiguity* is a binary variable, coded 1 if both states share a common boundary and 0 if they do not. *Major Power* is another binary variable, coded 1 if either one or both states in the dyad are a major power and 0 if neither are super powers. It is worth stressing that differently from all the other variables, these two variables have a negative hypothesized relationship to peace (if there is a major power in the dyad or the states share common boundary the expectation is that the risk of conflict will increase, while the probability for peace decreases). As a result the value 0 represents the maximum value for these two variables, while 1 the minimum (it is important to bear this in mind in the following discussion on model interpretation). *Distance* is a variable that is measured as the logarithm to the base 10 of the distance in kilometers between the two states' capitals. *Capability* is the logarithm to the base 10 of the power ratio between the two states (measured on stronger country to weak country) with power calculated on the basis of three indicators: demographic, industrial, and military power. The demographic capability is equal to the total national population plus number of people in urban areas. Industrial power is given by industrial energy consumption plus iron and steel production. Military capability is measured as a number of military personnel in active duty plus national military expenditure over the last 5 years. The two final variables are *Democracy* and *Dependency*.

*Democracy* is measured on a 21 scale where 10 is the highest value and -10 is the lowest value and it represents the degree of democratization in the less democratic state in the dyads. The variable *Dependency* is a continuous variable measuring the level of economic interdependence of the less economically dependent state in the dyad. To measure interdependence in each country in the dyads the sum of the country's import and export with its dyadic partner is divided by the country's Gross Domestic Product (dyadic trade as a portion of a state's gross domestic product) and only the lowest interdependent value between the two states is included in the analysis. We lag all independent variables one year to make temporally plausible any inference of causation.

Our data set is the population of politically relevant dyads for the cold war and immediate post-cold war period (CW), from 1946 to 1992, as described extensively and used by Russett and Oneal[1]. This population is defined as all dyads containing a major power and/or all contiguous dyads. We chose the politically relevant population because it sets a hard test for our models. Omitting all distant dyads composed of weak states means we omit much of the influence which variables that are not very amendable to policy intervention (distance and national power) would exert in the full data set. By this omission we make our job harder by reducing the predictive power of such variables, but it also makes the exercise more interesting. By focusing only on dyads that either involve major powers or are contiguous, we test the discriminative power of the Bayesian neural networks on a difficult set of cases. The neural network is trained with only highly informative data since every dyad can be deemed to be at a risk of incurring a dispute, yet it is harder for the network to discriminate between disputes and non-disputes because the politically relevant group is more homogeneous (e.g., closer, more inter-dependent) than the all-dyad data set.

The unit of analysis is the dyad-year. There are a total of 27,737 cases in the cold war population, with 26,845 non-dispute dyad-years and 892 dispute dyad-years. In relation to the dispute dyad years only the initial years of the militarized disputes are included, since our concern is to predict the onset of a conflict rather than its continuation.

The dataset of the politically relevant dyads is used to generate two different sets, training and testing sets, with the training set used only for training and the test set to assess out-of sample accuracy. The validation set for training is not used because we are pursuing a Bayesian approach to neural network training, which does not over-fit the model and the dataset is, therefore, only divided into training and testing sets. The size of the training sets consists of 500 dispute and 500 non-dispute dyad-years, while the test data consists of 392 dispute and 26345 non-dispute dyad-years. The reason why the data is split in this way is because we are trying to address the rare-event prediction problems often associated with neural networks. The balanced training data has been created in order to give conflict and peace cases equal importance in training. The network performance is then evaluated in terms of the prediction based on the expected unbalanced set.

*Neural Networks*

In this study, neural networks that are formulated in the Bayesian framework and were trained using the evidence framework, based on Gaussian approximation and hybrid Monte Carlo methods[15] were used for dispute classification in conflict analysis. This section gives the over-view of neural networks within the context of classification problems. In this paper, a multi-layer perceptron (MLP)[16] adopting supervised learning was used to map the 7 dyadic variables ($x$) and the MID ($y$). The relationship between the $k^{th}$ MID, $y_k$, and the dyadic variables, $x$, may be written as follows [16]

$$y_k = f_{outer}\left（\sum_{j=1}^{M} w_{kj}^{(2)} f_{inner}\left(\sum_{i=1}^{d} w_{ji}^{(1)} x_i + w_{j0}^{(1)}\right) + w_{k0}^{(2)}\right) \quad (1)$$

Here, $w_{ji}^{(1)}$ and $w_{ji}^{(2)}$ indicate the weights in the first and second layers, respectively, going from input *i* to hidden unit *j*, *M* is the number of hidden units, *d* is the number of output units while $w_{j0}^{(1)}$ indicates the bias for the hidden unit *j* and $\omega_{k0}^{(2)}$ indicates the bias for the output unit *k*.

Selecting the appropriate network architecture is an important part of model building. In this paper, the architecture chosen was the MLP trained using the scaled conjugate gradient method[17]. In addition to selecting the MLP model, another important decision for model building lies in the selection of the right number of hidden units (*M*) and the type of functional transformations that they perform. This is because a higher number of *M* will produce highly flexible networks, which may learn not only the data structure but also the underlying noise in the data. Instead, too few hidden neurons will produce networks that are unable to model complex relationships. In this paper, the architecture that includes *M* and activation functions was chosen using Genetic Algorithms (GA) [18]. GA was inspired by Darwin's theory of natural evolution. Here, we use this natural optimization method to optimize the architecture of the MLP neural networks in the light of the data. To identify the optimal MLP architecture, the network was trained several times using scaled conjugate gradient method and GA used to perform a global search in the solution space and select the best solution[18]. The genetic algorithm in this paper used a population of binary-string chromosomes[18], representing different MLP architectures, discretized in the search space with a fitness function. Each time a network architecture is trained a fitness value is assigned to it. Then the GA selects the architectures with the highest fitness values and combining them to produce new solutions (a new population of chromosomes). GA was implemented in this paper through performing: (1) simple crossover; (2) binary mutation; and (3) roulette wheel reproduction. The details of these may be found in [19]. Four activation functions were considered. These were linear, logistic, hyperbolic tangent and soft-max [18] and *M* was restricted to be maximum $\frac{n}{r(i+o)}$ with *n* being the number of samples in the training, *i* and *o* representing the number of neurons in the input and output layer, respectively, and *r* being a constant set by the noise level of the data. GA population of 20 was used. The GA identified $M = 10$, logistic function in the output layer and the hyperbolic function in the hidden layers as the optimal architecture.

The problem of identifying the weights and biases in neural networks may be posed in the Bayesian framework as [16]:

$$P(w|D) = \frac{P(D|w)P(w)}{P(D)} \qquad (2)$$

where $P(w)$ is the probability distribution function of the weight-space in the absence of any data, also known as the prior distribution function and $D \equiv (y_1,...,y_N)$ is a matrix containing the MID data. The quantity $P(w|D)$ is the posterior probability distribution function after the data have been seen, $P(D|w)$ is the likelihood function and $P(D)$ is the normalization function also known as the "evidence". For the MLP equation 2 may be expanded using the cross-entropy error function to give [16]:

$$P(w|D) = \frac{1}{Z_s} \exp\left( \beta \sum_n^N \sum_k^K \{t_{nk} \ln(y_{nk}) + (1-t_{nk})\ln(1-y_{nk})\} - \sum_j^W \frac{\alpha_j}{2} w_j^2 \right) \qquad (3)$$

where

$$Z_s(\alpha,\beta) = \left(\frac{2\pi}{\beta}\right)^{N/2} + \left(\frac{2\pi}{\alpha}\right)^{W/2} \qquad (4)$$

The cost-entropy function was used because of its classification advantages [18] and the weight-decay for the prior distribution was assumed because it penalises weights of large magnitudes. In equation 3, $n$ is the index for the training pattern, hyperparameter $\beta$ is the data contribution to the error, $k$ is the index for the output units, $t_{nk}$ is the target output corresponding to the $n^{th}$ training pattern and $k^{th}$ output unit and $y_{nk}$ is the corresponding predicted output. The parameter $\alpha_j$ is another hyperparameter, which determines the relative contribution of the regularisation term on the training error. In equation 3, the hyperparameters may be set for groups of weights.

Equation 3 can be solved in two ways: by using Taylor expansion and approximating it by a Gaussian distribution and applying the evidence framework [15]; or by numerically sampling the posterior probability using Monte Carlo method [12]. In this paper both approaches are followed and the two formulations are compared in the context of the conflict modelling problem.

*Gaussian Approximation*

Bayesian training of MLP neural networks is essentially about the calculation of the distribution indicated in equation 3. One method of achieving this goal is to assume a Gaussian approximation of the posterior probability by Taylor expansion. If this assumption is made, the posterior probability of the MIDs can be calculated by maximizing the evidence [15]. The evidence is the denominator in equation 2. The evidence framework finds the values of the hyperparameters that are most likely, and then integrates them over the weights using an approximation around the most probable weights. The resulting evidence is then maximized over the hyperparameters. The evidence framework is implemented by following these steps:

(1) Infer the parameters $w$ for a given value of $\alpha$. This is calculated in this paper by using the scaled conjugate gradient optimization method [15];

(2) Infer the value of $\alpha$ by approximating equation 3 with a Gaussian distribution and maximizing the evidence given the most likely weights.

*Hybrid Monte Carlo (HMC) method*

Another way of solving equation 3 is by using the HMC method [12]. The main idea of this method is solve the problem by calculating the mean of a function, $f(w)$, sampled from the posterior distribution of the weights. Provided that the number of samples approaches infinity, this method is exact. The HMC method uses the gradient of the neural network error to ensure that the simulation samples throughout regions of higher probabilities. This causes the HMC to avoid the random walk associated with traditional Monte

Carlo methods. The details of the HMC are in Neal [12]. As a result, the HMC performs better than the traditional Monte Carlo method. The gradient is calculated using the backpropagation method [16]. Sampling using the HMC is conducted by taking a series of trajectories and then either accepting or rejecting a resulting state at the end of each trajectory. Each state is represented by the network weights and its associated momentum, $p_i$. Each trajectory is achieved by following a series of leapfrog steps, which are described in detail by Neal [12]. For a given leapfrog step size, $\varepsilon_0$, and the number of leapfrog steps, $L$, the dynamic transition between two states of the HMC procedure is conducted as follows:

(1) Randomly choose the direction of the trajectory, $\lambda$, to be either –1 for backward trajectory and +1 for forward trajectory;

(2) Starting from the initial state, *(w,p)*, perform *L* leapfrog steps with predefined step size resulting in state *($w_{new}$,$p_{new}$)*. Here *p* is the momentum vector and is described in detail in Neal [12], $\varepsilon_0$ is a chosen fixed step size and *k* is the number chosen from a uniform distribution and lies between 0 and 1;

(3) Reject or accept *($w_{new}$,$p_{new}$)* using the Metropolis criterion [19].

In the Metropolis criterion, if the current posterior probability given the weights and the data is higher than the previous posterior probability then accept the new sample, otherwise, accept it with a probability of *exp -(dE/T)* where *dE* is the change in error between the current sample and the previous sample.

*Neural Network Results*

Neural networks methods were implemented and the classification of conflict results obtained. To assess the performance of the classification results, the Receiver Operating Characteristics (ROC) graphs were used[20] and their results are shown in Figure 1.

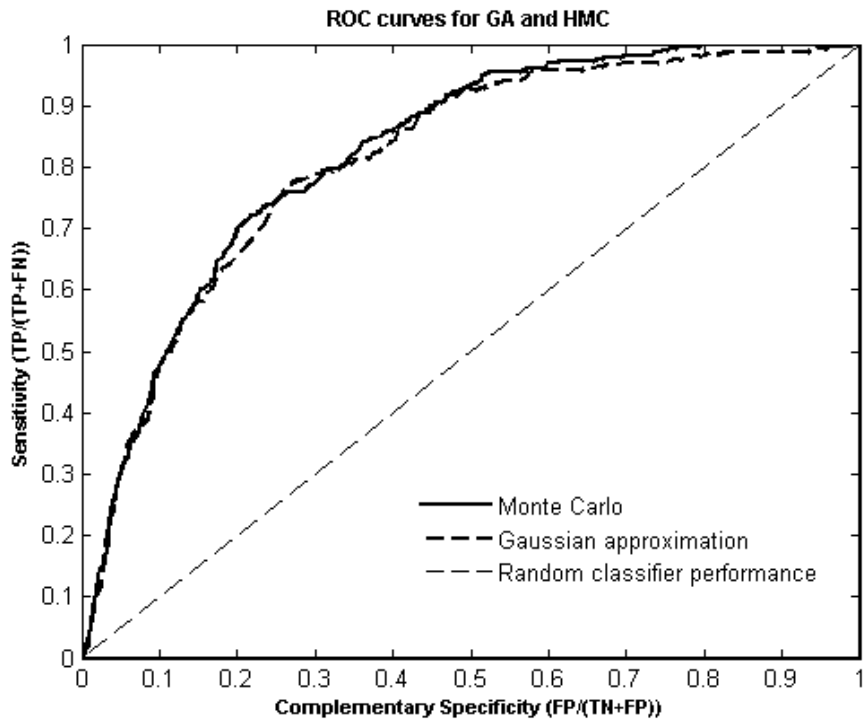

**Figure 1** The ROC of the classification of militarized interstate disputes

This measurement of performance was chosen since the ROC curves investigate the tradeoffs between false-positive and false negative for a variety of predictive thresholds, and do not penalize models whose prediction is biased too high or too low. In the ROC curves, the x-axis gives the proportion of disputes correctly predicted, while the y-axis provides the proportion of non-disputes correctly predicted for different thresholds. The general idea is that any threshold used as cut-off value between disputes and non-disputes will correspond to a single point on the ROC curve. The area under the ROC curve indicates how good the classifier is. If the area under the ROC curve is 1 then the classifier has classified all cases correctly while if it is 0 then the classifier has classified all cases incorrectly. In the ROC curve in Figure 1, both the classifiers give areas under a ROC curve of 0.82 which is a good classification rate. These results indicate that both the Gaussian and the HMC approach give the same level of classification accuracy.

When a confusion matrix was used to analyze the classification results of the two Bayesian methods, the results in Table 1 were obtained.

Table 1: Classification Results

| Method | TC | FP | TP | FC |
|---|---|---|---|---|
| Gaussian Approximation | 278 | 114 | 19462 | 6883 |
| Hybrid Monte Carlo | 286 | 106 | 19494 | 6851 |

TC = true conflict, FC = false conflict, TP = true peace, and FP = false peace

The confusion matrix contains information about actual and predicted classifications given by a classification system. When the accuracies of the two methods were calculated on the basis of the true positive rate (the proportion of disputes that are correctly identified), the HMC performs marginally better than the GA. As shown in Table 1, the HMC provides a true positive rate of 73%, while the Gaussian approximation gives a rate of 71%. In relation to the true negative rate (the proportion of non-disputes that are classified correctly) both methods perform the same with 74% accuracy.

On the basis of these results the HMC emerges as being marginally more accurate than the Gaussian approximation. This is primarily due to the fact that the Gaussian approximation is generally not as valid as the Monte Carlo approach. The reasons for HMC performing marginally better than GA can be seen by comparing two literature texts by Neal[12] and MacKay[10]. The implementation of Bayesian neural networks requires integration over high dimensional weight spaces. In order to make these integration steps more tractable, approximations are introduced which simplify their computation. Provided that the assumptions made hold the computation of the integrals will be fairly accurate. However, the use of Markov Chain Monte Carlo methods proposed by Neal[12] does not require any approximations and is therefore more general than the GA method. It is also worth noting that since the HMC is better at predicting disputes, this marginal improvement is quite important from a policy point of view. The HMC model identifies 8 more politically costly and dangerous dispute dyads than the Gaussian approach, therefore allowing 8 more disputes cases to be controlled and possibly be solved before they actually occur. Furthermore, this is done while reducing the rate of false positives, the proportion of disputes wrongly classified. HMC reduces the false positive of 32 cases compared to Gaussian approximation. This is an important result in a control environment since HMC allows better prediction on conflicts without wasting valuable resources on controlling disputes that are unlikely to happen. These results also stress that, in order to further improve accuracy, attention may need to be paid to data sources and variable construction.

*Model Interpretation and the Automatic Relevance Determination (ARD) Approach*

We now interpret the causal model, given by the neural networks developed in this paper. The interpretation of the causal hypotheses represented by a trained neural network is a complex exercise for several reasons. First, neural network models encode their knowledge across hundreds or thousands of parameters (weights) in a distributed manner. These parameters embed the relationships between the input variables and the dependent output. The sheer number of parameters and their distributed structure make the task of extracting knowledge from the network a difficult one. Second, the weight parameters of a multi-layer neural network usually represent non-linear and non-monotonic relationships across the variables, making it difficult to understand both the relative contributions of each single variable and their dependencies.

In order to understand both the relative and dependent contribution of each input on the MIDs, we utilize an indirect method and then the ARD approach. The indirect method provides an indication of the interactions existing among the inputs, while the ARD identifies the relevance of each input.

In the indirect method, we compare the model output with a new output produced by a modified form of the input pattern. When analyzing the causal relationships between input and output variables, the neural network shows that when the *Democracy* variable is increased from a minimum to a maximum, while the remaining variables are set to a minimum, then the outcome moves from conflict to peace. This is an indication that although interactions exist, *Democracy* also exerts a direct influence on peace. When all the variables were set to a maximum then the outcome was peace. When all the parameters were set to a minimum then the possibility of conflicts was 52% (is this value still the same with the new analysis?). These results are quite expected and indicate that all the inputs are quite important. When one of the variables was set to a minimum and the rest set to a maximum, then it was observed that the outcome was always peace. When each variable was set to a maximum and the remaining variables set to a minimum then the outcome was always conflict, with the exception of *Democracy* and *Dependency* where the outcome was peace. The first result stresses that strong interactions exist in relation to dispute patterns since no single low value can produce a dispute outcome. The second result indicates that more additive relationships than interactive ones are in place for peaceful patterns since one single maximum value for *Democracy* or *Dependency* can maintain peace. These results support recent findings by Lagazio and Russett[4], but also reveal new insights. *Democracy* and *Dependency* emerge as having a strong additive impact on peace. This means that these two variables alone could contribute significantly to peace, even without the positive influence of the others.

This section introduces and implements the ARD to understand the influence of the input parameters on the MIDs. The ARD model[10] is a Bayesian model that is used to determine the relevance of each input on the output. The ARD is constructed by assigning a different hyperparameter to each input variable and estimating the hyperparameters using a Bayesian framework. The input weights that have higher hyperparameters are not influential and have less effect on the output than the ones with lower hyperparameters .

In this paper, the ARD is used to rank the 7 variables used in the analysis with regards to their relative influence on the MIDs. The ARD was implemented, the hyperparameters calculated and then the inverse of the hyperparameters was calculated and the results are in Figure 2.

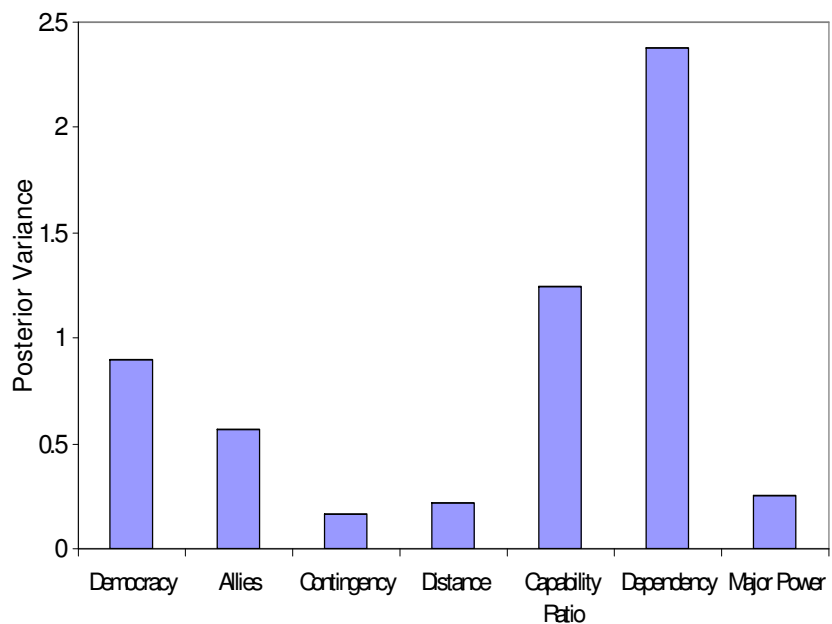

**Figure 2** A graph showing the relevance of each variable with regards to the classification of MIDs.

Figure 2 indicates that *Dependency* has the highest influence, followed by *Capability*, *Democracy* and then *Allies*. The remaining three variables, i.e. *Contiguity, Distance* and *Major Power,* have similar impact although it is smaller in comparison with other four variables. The results in Figure 2 indicate that the two variables, *Democracy* and *Dependency*, have a strong impact on conflict and peace outcomes. However, *Capability* and *Allies* cannot be ignored. Once again this confirms recent positions, which see the *Capability* and *Allies* as mediating the influence of *Democracy* and *Dependency* by providing constraints or opportunities for state action[3,4]. This means that a relatively equal dyadic power ratio, closer geographic proximity, and no alliance greatly increase the impact of a low level of economic interdependence and democracy on the probability of a dispute. The results of the ARD are further confirmed by analyzing the accuracy of the model as each of the inputs is omitted from the input. Figure 3 shows the ROC curves when each of the inputs is omitted, one at a time.

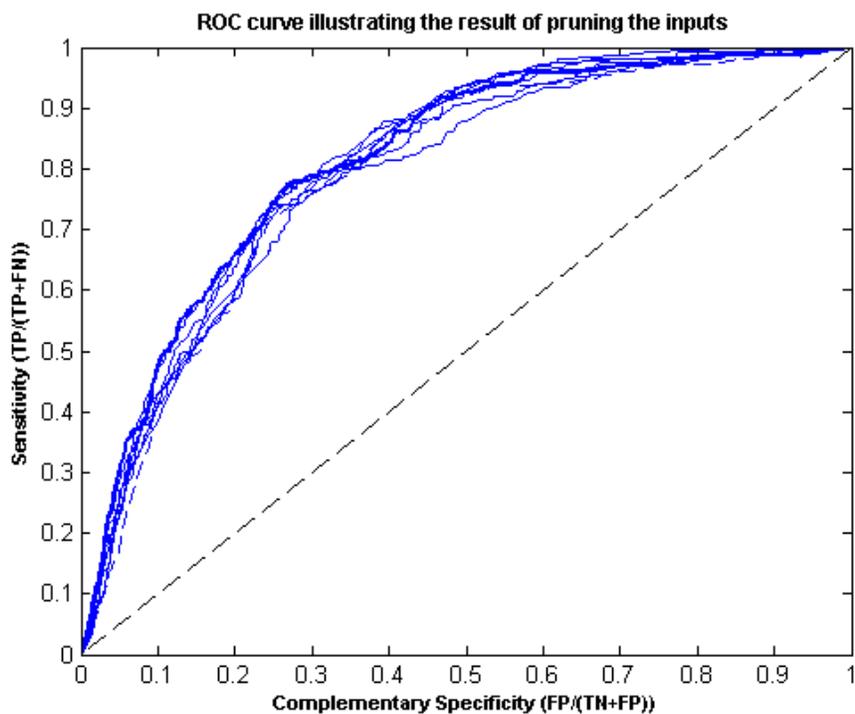

**Figure 3** The ROC curves when each of the inputs are left out

From looking at the curves it is obvious that by omitting just one of the variables irrespective of the importance, good classification can still be obtained. However, by analyzing the area under each of the curves, the results obtained from the ARD can be confirmed. Table 2 shows the AUC values when one of the inputs is removed.

Table 2: Input pruning

| input variable | Area under curve |
| --- | --- |
| Dependency | 0.7914 |
| Capability | 0.7945 |
| Democracy | 0.7951 |
| Allies | 0.8030 |
| Distance | 0.8080 |
| Major Power | 0.8173 |
| Contiguity | 0.8140 |
| None | 0.82 |

Figure 4 compares the classification performance of the four most significant variables versus the three less significant variables.

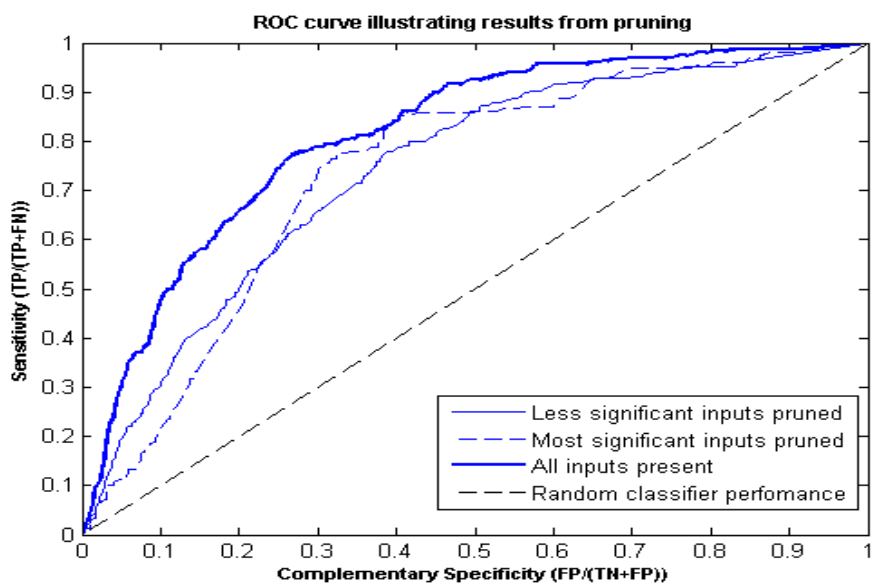

**Figure 4** The ROC curves showing the performance of the 4 most significant variables versus the 3 less significant varialbles.

It is obvious from looking at the ROC curves that although contiguity, distance and major power are the least significant variables, they are necessary to the performance of the neural network model. This is because the ROC curve of these variables gives better accuracy than the four significant variables in some regions.

Overall, our results support theories of the liberal peace identifying both democracy and economic interdependence as key variables in relation to peace and war[5]. Furthermore, they confirm that strong interactions exist between the democracy and dependency and capability and allies in relation to dispute patterns. Therefore, relationships across the variables do appear to be highly non-linear and contingent. Low economic interdependence and democracy play important direct as well as indirect roles in producing war. Their influence is greatly strengthened by their interactions with capability and alliances.

**4. Control of conflict**

Now that we have developed a model to predict the MIDs, given dyadic variables, the next step is to use this methodology to identify the set of variables that ensure that conflict can be controlled, thus reducing the probability of occurrence of war in the international context. The whole rationale behind the development of the interstate dispute prediction infrastructure is to maximize the occurrence of peace, while minimizing conflicts. This is achieved in this paper by applying classical control theory to conflict resolution. Classical control theory has been used to control many complex problems. A literature review on the application of control system, to solving complex problems, can be found in[21]. In this article Shurgel reviews recent developments of bioprocess engineering that include the monitoring of the product formation processes. He also reviews the advanced control of indirectly evaluated process variables by means of state estimation using structured and hybrid models, expert systems and pattern recognition for process optimization. Control system theory has also been applied to aerospace engineering, where it has been applied to actively control the pressure oscillations in combustion chambers[22]. Genetic algorithms and fuzzy logic have been successfully used to control the load frequency in PI controllers[23]. Plant growth has been optimally controlled using neural networks and genetic algorithms[24] and fuzzy control has been used for active management of queuing problem[25]. Other applications of control methods to complex systems may be found in[27-28].

In this paper, we use control system theory to control interstate conflict. This is done by identifying controllable variables that will produce a peaceful outcome. To achieve this, the cost function is defined as the absolute value of the neural network prediction, which should be as close as possible to zero, i.e. absolute peace. Two approaches are used: a single strategy approach where only one controllable variable is used and a multiple strategy where all the controllable variables are used. Of the 7 dyadic variables discussed earlier in the paper, there are only 4 that are controllable and these are: *Democracy*, *Allies*, *Capability* and *Dependency*. Therefore, only these variables will be part of the control analysis.

In this paper, the control system infrastructure consists of three components: the HMC trained Bayesian feed-forward neural network developed in previous sections, which predicts the MIDs, as well as the optimizer, which is activated only if the predicted outcome is dispute, and therefore undesirable, and whose function is to identify the controllable input parameters that could produce peace. This approach is illustrated in Figure 5.

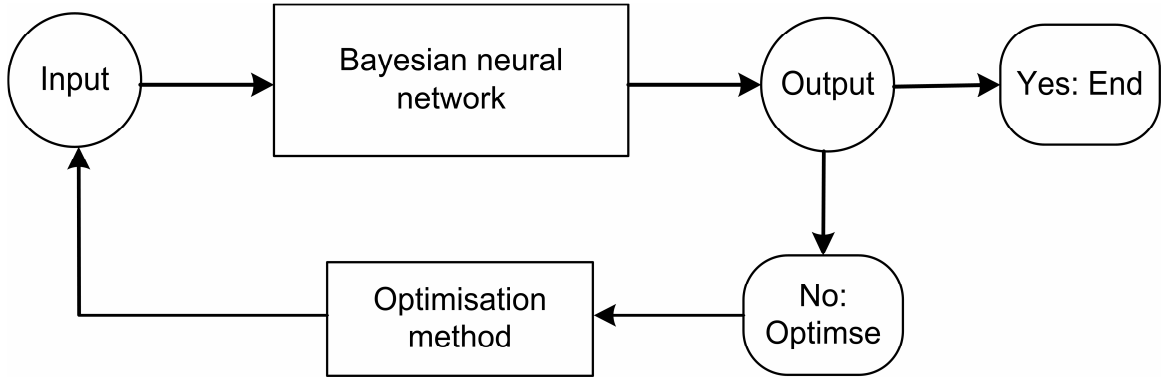

**Figure 5** Feedback control loop that uses Bayesian neural networks and optimization method

The optimizer can be any nonlinear function minimization method. In this study the Golden Section Search (GSS) method[29] was used for single strategy approach and Simulated Annealing (SA)[30] was used for the multiple strategy approach. The use of the GSS method is primarily because of its computational efficiency. It should be noted here that other methods, such as the conjugate gradient method, scaled conjugate method or genetic algorithm might also be used[29].

On implementing the control strategy the Bayesian network using HMC for training was used. This is because of its better performance on dispute prediction. The control approach was implemented to achieve peace for the dispute data in the test set. There were 392 conflict outcomes in the test set of which 286 were classified correctly by the HMC (see Table 1). Therefore, in this paper, we control the 286 dispute cases by identifying the controllable variables that will produce peace. Of course in a real application we would not know which predicted conflicts are true and which are false because it is in the feature, therefore we would apply the control approach to all predicted conflict cases. However, this could be problematic since the HMC model still predicts too many false conflicts. As a result we would control for cases that need no control, wasting valuable resources. In order to avoid inefficiency and reduce the impact of false negative on the control component of the system, the confidence level for the predicted outcomes can then be used as further criteria to identify true predicted conflict cases. Since we found that a significant number of false conflicts predicted by the HMC model present low confidence level, the inefficiency of the control system can be readdressed by disregarding predicted conflict cases with low confidence.

When the control strategies were implemented, the results shown in Figure 6 were obtained. These results show that for a single strategy approach, where *Democracy* is the controlling variable, 90% of the 286 conflicts could have been avoided. When the controlling variable *Allies* is the only variable used to bring about peace it was found that 77% of 286 conflicts could have been avoided. When either *Dependency* or *Capability* was used as a single controlling variable, 100% of 286 conflicts could have been avoided. In relation to the multiple strategy approach, when all the controllable variables were used simultaneously to bring about peace, all 286 conflicts were avoided.

The results in Figure 6 show the original value of the dyadic variables in a specific dyadic conflict case as well as the new values suggested by the system in order to obtain peace using a single strategy method. The new value stresses how much each variable is singularly changed to give a peaceful outcome. In this case *Democracy* requires the smallest relative change while *Dependency* the highest. The result show that all the variables need to change less compared to the single approach in order to produce peace.

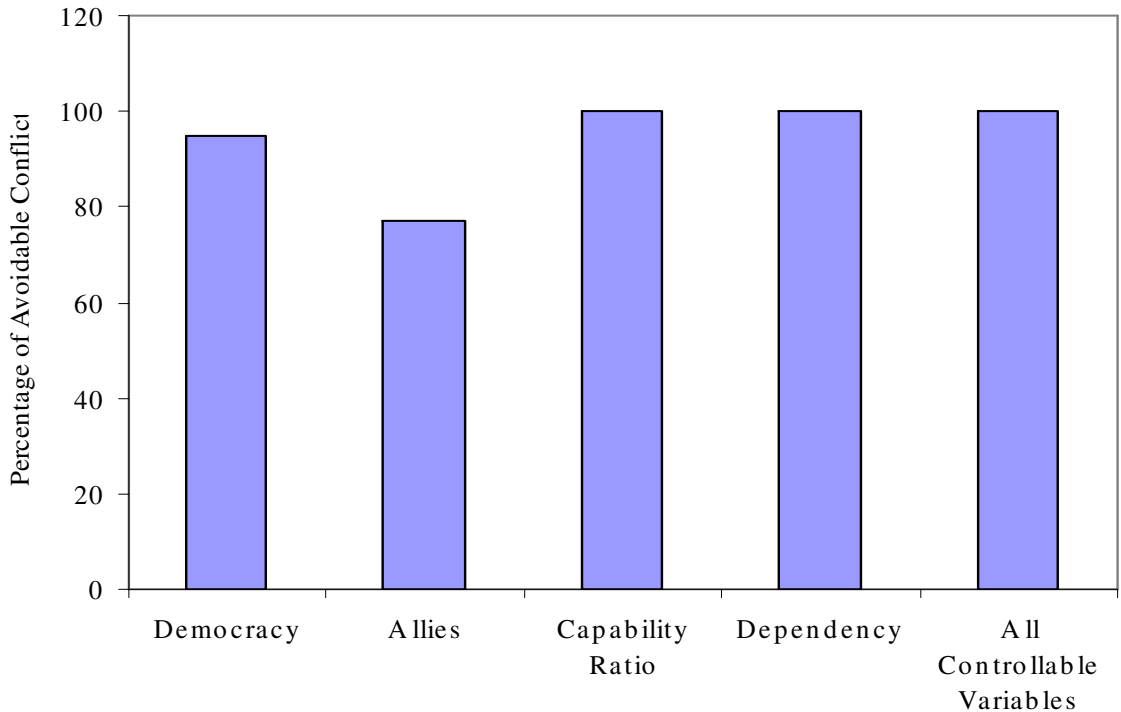

**Figure 6** Graph showing the proportion of past conflicts that could have been avoided.

This is expected since in this case all the variables have been modified simultaneously. However, the change required for *Democracy* and *Dependency* is only marginally different from the one required by these variables in the single strategy, while *Allies* and *Capability* need to change significantly less compared to the single approach. This is a further indication of the key importance of democracy and dependency in order to achieve peace. Although we change other variables, we still need to change democracy and dependency almost as much as in the single strategy in order to achieve peace. In addition this result indicates that, in this case, the single approach on key variables is more efficient than the multiple one since the required change for democracy and dependency is almost similar in both approaches but the single approach does not require further changes in the other variables.

On the basis of these results it is evident that the user could decide different strategies to obtain peace. The user (decision maker) should decide which approach, single or multiple, is the most convenient on the basis of how easy and cost efficient is the intervention on the identified controllable variables and how long it would take to control them. This could be done on case by case basis, where additional analyses, provided by human experts, could then identify the optimal control strategy out of those suggested by the control system approach. It is also worth noticing that the lowest relative change may not be the most convenient strategy to follow. This is because some variables may be more difficult to control than others in a specific context even if they require a small change. Other time the intervention on some variable can be more costly because of particular political conditions (for instance regime change in Afghanistan compared to regime change in Eastern European countries). Furthermore, some variables are less amenable to real change in the political system than others. For instance capability is a variable than can change only very slowly and on a timescale of decades since it takes time to alter demographic and industrial patterns. In comparison allies and dependency, followed by democracy, are more sensitive to change therefore intervention on these variables will require comparatively less time to produce the desire outcome. Finally in relation to how we could manipulate the variables, political scientists sometimes provide different solutions. *Allies* and *Dependency* are probably easier to manipulate. *Allies* can be controlled by opening up existing security cooperation agreements to new members and devising international policies to strengthen multinational cooperation, while *Dependency* can be controlled by boosting bilateral economic and trade relations (see example of China with the USA). The intervention on *Democracy* is more problematic since different approaches exist within democratization theory on how we could encourage democracy. Some authors focus on external factors and how international organizations or/and the international community could promote democratization through peace building initiatives, support of democratic movements, foreign aid in exchange of good governance and transparency, embargo, and even military intervention. Others underline the importance of internal factors and how some internal antecedent conditions, such as economic development need to be in place in order to trigger democracy. However democracy may work in some states with low economic development and it may be cheaper and easier to achieve an initial higher level of democratization than higher level of economic development (examples of this can be Cambodia and Nigeria). As we mentioned earlier *Capability* is not easy to manipulate since required boosting industrial and demographic development not only military expenditure. All these options need to be evaluated, case by case, since some interventions could be feasible in the same dyad but not in other. As a result, the control approaches suggested by the system, must be conjoined with political practicalities and integrated by analyses provided by human users (decision makers), selecting the optimal control strategy out of the ones suggested by the system. Therefore, human analysis on which specific policy option is the most practical in order to achieve the variable change suggested by the system is an important part of the final decision making process.

**5. Conclusion**

In this paper Bayesian neural networks were used to model the relationships between democracy, allies, contiguity, distance, capability as well as dependency and militarized disputes. Gaussian approximation and hybrid Monte Carlo (HMC) method were used to train the Bayesian neural networks and it was found that the HMC was marginally more accurate than the Gaussian approximation. The analysis of the influences of the input parameters on the militarized disputes was conducted through the indirect and automatic relevance determination (ARD) approaches. The results obtained indicated that strong interactive relationships exist among the variables and that *Dependency* carries the most weight, followed by *Capability*, then *Democracy* and then *Allies*. Finally two control approaches were implemented to identify control strategies for maximizing peace. The single strategy approach was implemented using the golden section search method, whereas simulated annealing was used for the multiple strategy approach. It was observed that all four controllable dyadic variables could be used simultaneously to avoid all correctly identified conflict. Furthermore, it was observed that either *Dependency* or *Capability* could also be used to avoid all the correctly predicted conflicts, followed by controlling only *Democracy* which results in 90% of disputes being avoided and then controlling *Allies* which results with 77% of disputes being avoided. Finally, by comparing findings from the single and multiple approach, it emerge that *Dependency* and *Democracy* are key variable to achieve peace since even in a multiple approach their required change is still quite significant and closed to their single level requirement in comparison to the other dyadic variables. This means that significant changes in *Dependence* and *Democracy* are necessary even if the other dyadic variables have been positively manipulated to achieve peace. It is also worth noticing that all the methods implemented in this paper to reveal the neural network model (indirect, ARD and control methods) provide similar findings, therefore identifying a similar causal structure among the variables. This is an important result since it underlines that neural network models can be effectively interpreted. It is worth noting that the methods used in this research can be used only as a suggestion as to direction that can be taken in foreign policy making. The neural network itself is a good model of complex input-output relationships but considered by many as non deterministic. It would not be advisable to apply this model blindly without checking whether the suggested solutions coincide with existing models. However, current research is directed into better understanding the neural network so that any control system using them can be analyzed quantitatively for properties such as stability.